\documentclass[RNAAS]{aastex62}

\newcommand\latex{La\TeX}

\begin{document}

\title{Prestige Bias on Time Allocation Committees?}


\correspondingauthor{J. S. Greaves}
\email{email greavesj1 at cardiff.ac.uk; ORCID 0000-0002-3133-413X}

\author{J. S. Greaves}
\affiliation{Cardiff University, School of Physics \& Astronomy, 4 The Parade, Cardiff CF24 3AA, UK}





\section{} 

Fairness is a key issue in the careers of astronomers \citep[e.g.]{caplar,massey}. I examine here the anecdotal suggestion that ``you're more likely to get time if you're on the TAC'', using public and published data for a large international telescope facility. This facility makes extensive efforts to appoint Time Allocation Committees (TACs) that are diverse across gender, region and career-point, factors seen to affect time awards \citep{spekkens,lonsdale,patat,reid}. The TAC also has a multi-panel structure so that  members are not rating each other's work face-to-face, mitigating against personal impact. The factor I consider here is different: since TAC appointment is regarded as an achievement, a ``prestige'' bias could operate, with other astronomers present at the meeting subconsciously viewed as ``{\it a good scientist, like myself}''. 

\section{Methodology}

I counted proposals awarded time for which the applicant was the Principal Investigator (PI). These values are N, as shown in Figure \ref{fig:1} (mean-N $<1$, individual N = 0-4, all per-round). A success {\it rate} could not be computed, as data for unsuccessful proposals and PIs are not public. N $\geq 1$ is simply a goal-measure (i.e. an applicant hopes to receive data that year). 

The TAC sample served on the committee within the period of rounds numbered 3-5, but not immediately before or after, i.e. not in any of rounds 1,2,6,7. Their mean-N were calculated before and after, and for while serving, separated by when individuals were on and off. The need to obtain a clean sample resulted in a moderate-sized group of 22 individuals (after dropping 3 persons as `unlikely to propose', i.e. not appearing as PI or co-I in any project in the database). The TAC-members served between 1 and 3 rounds each (average of 1.6), and it was assumed they were able to be a PI in all rounds. 

Control samples were selected randomly by letting the database query tool auto-complete for author surnames with various initial letters. The aim was to return persons associated with proposals but not necessarily as PI, and with similar-sized groups to the TAC sample. Control samples C1, C2 and C3 consist of 23, 29 and 26 non-TAC-serving individuals, respectively, and these were combined to give mean and standard deviation for N(C).

\section{Statistics}

\begin{figure}[h!]
\begin{center}
\includegraphics[scale=0.6,angle=0]{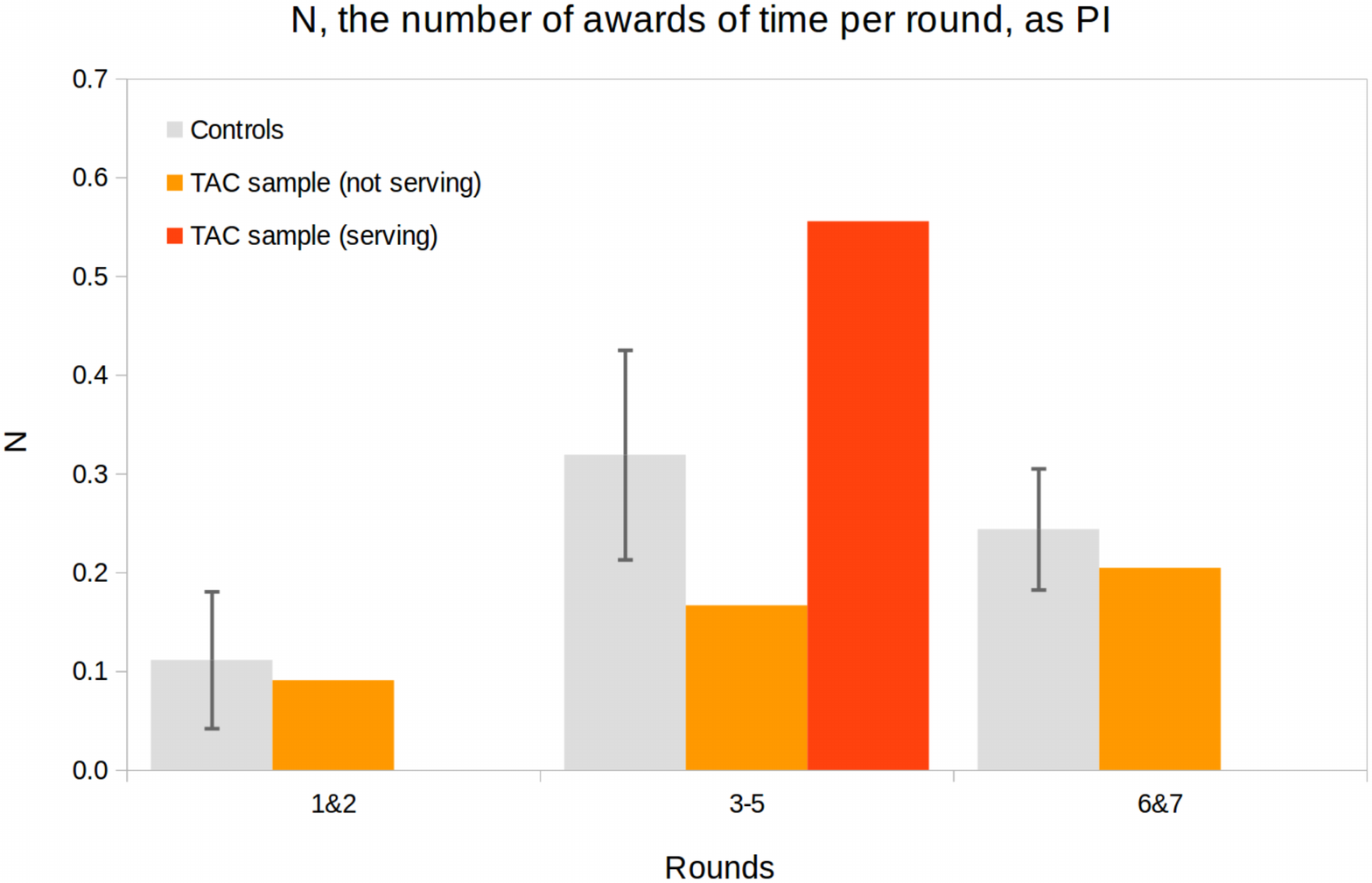}
\caption{N (proposals accepted per-round, as PI) for the TAC and control (C) samples. Grey bars are for C, with red and orange bars for TAC members when on and off the TAC, respectively. Errors for N(C) are the standard deviation of samples C1, C2, C3 in the same period. N(C) trends reflect facility supply-and-demand. N(round-7) will slightly increase as lower-priority projects get added to the schedule. Data are available at \href{http://journals.aas.org//authors/data.html\#DbF}{DataTable.txt}.\label{fig:1}}
\end{center}
\end{figure}

Figure \ref{fig:1} compares N for TAC-members to the controls, C. When not serving, TAC-members performed slightly {\it worse}, but only by -0.8$\sigma_{\rm C}$ on average. This difference is not significant, but could encompass small population differences. The TAC includes theorists, some of whom may never be involved in proposals (unlike the controls, found in the approved-project database). Further the TAC sample was unusual in being majority-female. An extensive gender check\footnote{Best-guesses were made by name/photo; no disrespect is intended to transgender and non-binary identitities.} of  successful round-5 PIs was combined with published numbers of female- and male-led proposals. For women, N was 0.03 below the mean\footnote{With a small $\Delta$N advantage, male-PIs had notably greater success ($\times$1.35) over female-PIs in getting {\it multiple} proposals accepted.}, which would drive N(TAC) downwards, but by only $\approx$0.01.  

The main result is that {\it N becomes considerably larger for TAC members at meetings they attend}. Compared to when off the TAC, serving improved success {\it three-fold} (N = 0.17 $\rightarrow$ 0.56, within the same period). After rolling off, success dropped by almost as much (N = 0.56 $\rightarrow$ 0.20). N(TAC) increased by 3.7$\sigma_{\rm C}$ between not-serving and serving (in the same time-period), even with $\sigma_{\rm C}$ probably over-estimating random errors (national differences were apparent between C-samples). The TAC-effect is dominant, e.g. boosting N ten times more than the +0.04 male-female difference. 

\section{Inferences}

The TAC-serving boost in getting proposals accepted supports the hypothesis of a bias in operation. The magnitude of the effect is remarkable, and not readily explained in a fair scenario. Alternate models include TAC-members being mainly driven to propose while serving, but this seems doubtful for applications to a highly-sought facility. Or, TAC-members could gain expertise at the meetings, and so start to apply more successfully (itself, perhaps ``unequal access'') -- but N(TAC) fell back substantially afterwards, contrary to such a model.  

The sample sizes here are constricted, as people roll on and off the TAC yearly, making it difficult to track uniform cohorts. A larger study could:-- follow every TAC-member over a larger number of rounds; investigate whether similar trends occur for other telescopes; and use non-public data to track success-probabilities as well as projects accepted. 

With these caveats on the dataset, the magnitude found here for the {\it ``more likely to get time if you're on the TAC''} effect is troubling. Careers are influenced by getting ground-breaking data, so a larger study of this skew in time-allocation is well-merited. The methodology used here attempted to remove other  trends (with time, gender, nationality) to leave the ``prestige bias'' of being a TAC-appointed scientist as the remaining factor, and the data support that this appears relevant. However, whether an unconscious or conscious reward-mechanism, or some other process, is at work is not clear.

\end{document}